\documentclass{article}

\usepackage{arxiv}
\usepackage[utf8]{inputenc}  
\usepackage{amsmath,amsfonts,amssymb}
\usepackage[T1]{fontenc}     

\usepackage{amsmath}
\usepackage{ragged2e}
\usepackage{booktabs}        
\usepackage{amsfonts}        
\usepackage{nicefrac}        
\usepackage{microtype}       
\usepackage{graphicx}
\graphicspath{ {./Figures/} }
\usepackage{float}
\usepackage{subfig}
\usepackage{enumitem}
\usepackage{longtable}
\usepackage[textwidth=6em,textsize=tiny]{todonotes}
\usepackage{color}
\usepackage{xcolor}		
\usepackage[colorlinks = true,
            linkcolor = purple,
            urlcolor  = blue,
            citecolor = cyan,
            anchorcolor = black]{hyperref}
\usepackage{soul}
\usepackage[numbers,square]{natbib}

\sffamily

\title{Segmentation-Free Outcome Prediction in Head and Neck Cancer: Deep Learning-based Feature Extraction from Multi-Angle Maximum Intensity Projections (MA-MIPs) of PET Images}

\author{
 Amirhosein Toosi \\
  Department of Radiology\\
  University of British Columbia\\
  Department of Integrative Oncology\\
  BC Cancer Research Institute\\
  Vancouver, BC \\
  \texttt{amirhosein.toosi@ubc.ca} \\
  \And
 Isaac Shiri \\
  Division of Nuclear Medicine and Molecular Imaging\\ 
  Geneva University Hospital\\
  Department of Cardiology \\
  University Hospital Bern \\
  Switzerland \\
  \texttt{isaac.shirilord@insel.ch} \\
  \And
 Habib Zaidi \\
  Division of Nuclear Medicine and Molecular Imaging\\ 
  Geneva University Hospital\\
  Geneva, Switzerland \\
  \texttt{habib.zaidi@unige.ch} \\
  \And
 Arman Rahmim \\
  Departments of Radiology and Physics\\
  University of British Columbia\\
  Department of Integrative Oncology\\
  BC Cancer Research Institute\\
  Vancouver, BC \\
  \texttt{arman.rahmim@ubc.ca} \\
}
\begin{document}
\textbf{\textit{This manuscript has been published in Cancers journal, Volume 16, Issue 14, P2538, Jul. 17, 2024}} \\

\url{https://doi.org/10.3390/cancers16142538}
\vspace{400pt}
\maketitle
\begin{abstract}
We introduce an innovative, simple, effective segmentation-free approach for outcome prediction in head \& neck cancer (HNC) patients. By harnessing deep learning-based feature extraction techniques and multi-angle maximum intensity projections (MA-MIPs) applied to Fluorodeoxyglucose Positron Emission Tomography (FDG-PET) volumes, our proposed method eliminates the need for manual segmentations of regions-of-interest (ROIs) such as primary tumors and involved lymph nodes.  Instead, a state-of-the-art object detection model is trained to perform automatic cropping of the head and neck region on the PET volumes. A pre-trained deep convolutional neural network backbone is then utilized to extract deep features from MA-MIPs obtained from 72 multi-angel axial rotations of the cropped PET volumes. These deep features extracted from multiple projection views of the PET volumes are then aggregated and fused, and employed to perform recurrence-free survival analysis on a cohort of 489 HNC patients. The proposed approach outperforms the best performing method on the target dataset for the task of recurrence-free survival analysis. By circumventing the manual delineation of the malignancies on the FDG PET-CT images, our approach eliminates the dependency on subjective interpretations and highly enhances the reproducibility of the proposed survival analysis method. The code developed for this work is publicly available at: \url{https://github.com/Amirhosein2c/SegFreeOP} .
\end{abstract}


\keywords{Artificial intelligence \and machine learning \and deep learning \and PET imaging \and head and neck cancer}

\section{Introduction}

Head and neck cancer (HNC) is the seventh most common cancer worldwide. It encompasses malignancies involving the anatomic sites that compose the upper aerodigestive tract, including the oral cavity, pharynx, larynx, nasal cavity, paranasal sinuses, and salivary glands \cite{Gormley2022-yc}. In 2016, over 1.1 million new cases and 4.1 million prevalent cases were reported, resulting in approximately 500,000 deaths worldwide \cite{mody2021head}. Although the median diagnosis age for HNC is around 60 years, in recent years the incidence of these cancers has increased in adults younger than 45 years, mainly due to higher numbers of oropharyngeal cancers associated with oncogenic human papillomavirus (HPV) \cite{marur2010hpv} .


Accurate prognosis and staging is crucial for improving patient care, treatment planning, and disease management, ultimately leading to improved survival rates \cite{Bossi2019-vs}. Recurrence-free survival (RFS) analysis is a common outcome prediction method used to assess the effectiveness of treatments for HNC. RFS measures the length of time after treatment during which the cancer does not recur. RFS plays a vital role in the management of HNC patients by providing information on the durability of treatment effects and helps to determine the optimal treatment strategy for individual patients \cite{Hashmi2020-qd}. RFS analysis is typically performed utilizing clinical and pathological features like tumor stage, grade, histological sub-type, and nodal involvement. However, relying solely on these factors may not fully capture the complex tumor heterogeneity and dynamic changes in the tumor micro-environment, which can limit the accuracy of prognostication \cite{Woolgar2006-gm}. 

Biomedical imaging methods such as computed tomography (CT), magnetic resonance imaging (MRI), and positron emission tomography (PET) can provide valuable information in various steps of HNC management \cite{Economopoulou2019-jr}. Specifically, PET imaging plays an active role in staging, treatment planning, monitoring response, recurrence detection and surveillance for HNC patients \cite{al2009clinical, castaldi2013role}. Several quantitative parameters could be extracted from fluorodeoxyglucose (FDG) positron emission tomography (PET) and CT images, providing information about the metabolic activity and aggressiveness of the tumor. This information can be helpful in predicting treatment response, disease progression, and patient survival. Some commonly used imaging biomarkers are Standardized Uptake Value (SUV) thresholds, such as SUVmean, SUVmax, SUVpeak, volumetric measures, such as Metabolic Tumor Volume (MTV) and Total Lesion Glycolysis (TLG) \cite{differding2015pet}. Additionally, the high-throughput extraction of radiomics features from PET-CT image modalities provides massive amount of hidden information that could help personalized management of the HNC \cite{vallieres2017radiomics, wong2016radiomics}. 

Imaging biomarkers and radiomics features are tumor specific; i.e. they need to be extracted from the tumor region of intertest (ROI). Therefore, accurate manual segmentation of ROIs within medical images is a crucial step for extracting these features. Manual segmentation has been the fundamental step in various medical imaging analysis tasks \cite{Kevin_Zhou2019-zw, Zanaty2016-pl}. However, it presents several challenges that can impact the accuracy, reproducibility, and efficiency of the analysis. Some of the key challenges of manual segmentation can be categorized as follows:

\textbf{Subjectivity and Inter-observer Variability:} Manual segmentation is a subjective process that relies on the expertise and judgment of the human annotators. Different individuals may interpret the image data differently, leading to inconsistencies in ROI delineation. Inter-observer variability, where different annotators produce different segmentations for the same image, can significantly affect the reliability and generalizability of the analysis \cite{Davico2023-vh}.

\textbf{Time and Labor Intensive:} Manual segmentation is a time-consuming task that requires significant human effort and expertise. The process of precisely outlining complex anatomical structures and tumor regions can be tedious and prone to errors \cite{Rehman2023-ca}. Moreover, the availability of skilled personnel for manual segmentation may be limited, especially in high-volume clinical settings, potentially causing delays in the analysis pipeline \cite{Iqbal2023-ae}.

\textbf{Limited Scalability and Generalizability: }Manual segmentation is not easily scalable to large datasets or multiple centers due to its labor-intensive nature. Generating manual segmentations for a large cohort of patients can be impractical and time-prohibitive \cite{Alba2018-fh}. Additionally, the segmentation outcomes may be influenced by the experience and training of the annotators, limiting the generalizability of the analysis across different sites and populations \cite{cardinell2022investigating, Rajiah2014-rz}.

\textbf{Variability in Tumor Characteristics:} Tumors in HNC can exhibit diverse characteristics, including irregular shapes, heterogeneous intensity patterns, and variable contrast uptake \cite{Johnson2020-ak}. Accurate delineation of these tumors solely based on visual cues can be challenging, leading to uncertainties and errors in manual segmentation \cite{Oreiller2022-ul}. This variability in tumor characteristics further complicates the segmentation process and introduces additional sources of variability and bias.

\textbf{Limited Representation of Tumor Heterogeneity:} Manual segmentation often focuses on delineating the tumor or the region of interest as a whole, neglecting potential subregions within the tumor that exhibit distinct characteristics \cite{John2019-rb, Rathore2018-yq}. This limited representation of tumor heterogeneity may overlook critical information relevant to recurrence-free survival prediction, as various subregions might contribute differently to the prognosis.

To address these challenges, we propose a segmentation-free approach for recurrence-free survival analysis in HNC. By circumventing the manual segmentation step, our approach eliminates the dependency on subjective interpretations and enhances the reproducibility of the analysis. Instead, our method utilizes a deep learning-based object detection model to automatically crop the head \& neck region in the FDG PET volumes, enabling a more efficient, objective, and reproducible analysis.

In the subsequent sections, we describe the methodology employed to automatically crop the PET volumes (sections \ref{subsec:method_overview} to \ref{subsec:od_training}), generate multi-angel maximum intensity projections of the cropped PET volumes (section \ref{subsec:MA-MIP}), extract deep features using a pre-trained deep convolutional neural network (sections \ref{subsec:feat_ext} to \ref{subsec:feat_fusion}), and perform recurrence-free survival analysis (section \ref{subsec:ML_pipeline}). We present the results of our study in section \ref{sec:results}, discuss their implications, and highlight the advantages of the segmentation-less approach in overcoming the challenges associated with manual segmentation in medical imaging analysis in section \ref{sec:discussion}.

\section{Materials and Methods}
\label{sec:materials_methods}
\subsection{Dataset}
\label{subsec:dataset}

For this work we used the training dataset of the third MICCAI Head and Neck Tumor segmentation and outcome prediction (HECKTOR2022) challenge \cite{Oreiller2022-ul, oreiller2022head, andrearczyk2022head}. The dataset comprises FDG PET/CT images, along with the clinical and survival data for a cohort of 489 patients with histologically proven H\&N cancer located in the oropharynx region. Scans were gathered across 7 different medical centers, with variations in the scanner manufacturers and acquisition protocols \cite{Andrearczyk2023-jz}. All 489 patients underwent radiotherapy treatment planning and had complete responses to the treatment (i.e. disappearance of all signs of local, regional and distant lesions). 
CT images had an original median voxel size of ${0.98 \times 0.98 \times 2.80}$ ${mm}^3$, and the PET images had a median voxel size of ${4.00 \times 4.00 \times 3.27}$ ${mm}^3$.
To ensure consistency, for all PET/CT images, both CT and PET volumes were resampled to a voxel size of ${1.0 \times 1.0 \times 1.0}$ ${mm}^3$ using a third-order spline method.

\subsection{Method Overview}
\label{subsec:method_overview}
\begin{figure}[h]
	\centering    
    \includegraphics[clip,width=1\linewidth,trim=6.75cm 2.8cm 8.5cm 2.8cm]{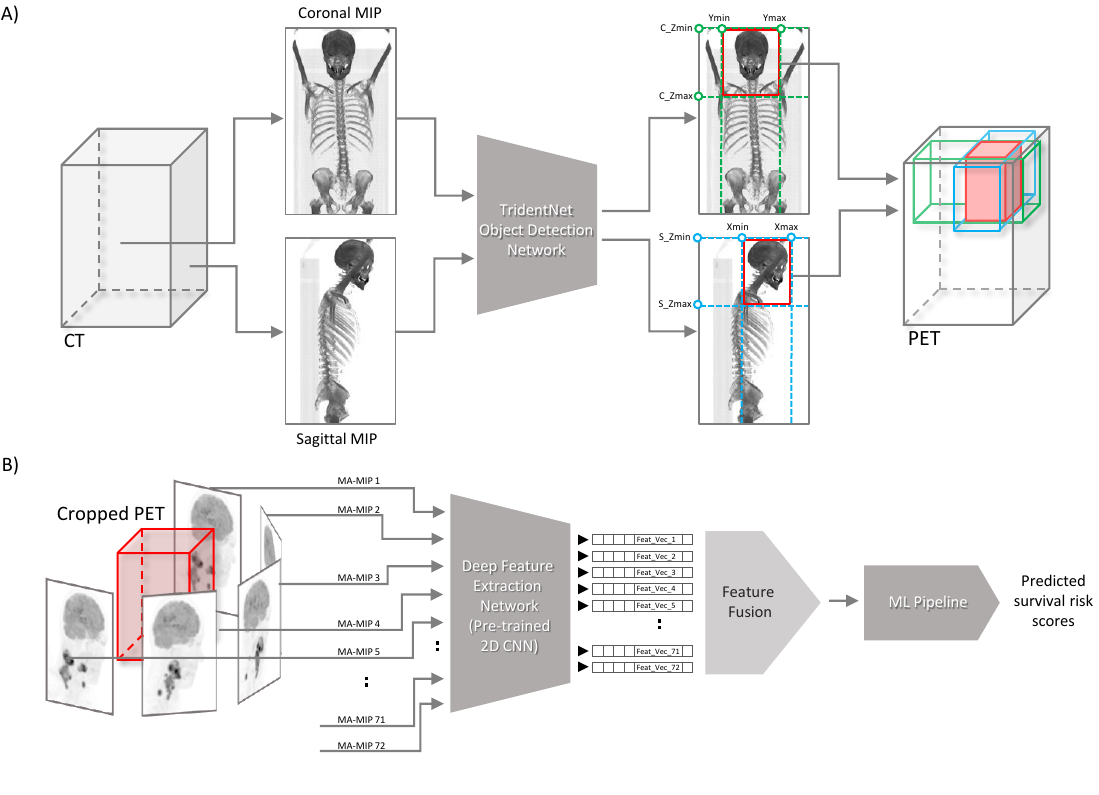}
\vspace{-2.5em}
	\caption{Overview of the proposed method for fully automated segmentation-free outcome prediction for head and neck cancer patients. A) Overall representation of the automated head and neck region detection and cropping. B) Extracting deep features from MA-MIP projections followed by feature fusion step.}
	\label{fig:Fig1}
\end{figure}

Here we describe our proposed segmentation-free outcome prediction method in detail (see figure ~\ref{fig:Fig1}). First, we train the object detection model \cite{Li2019-ge} on manually drawn bounding boxes containing the anatomical region of patient's head and neck, on both coronal and sagittal maximum intensity projections of CT images, as described in section \ref{subsec:od_training}. The trained network is then used for cropping the head and neck region of the corresponding registered FDG PET image volume of the patient, as described in section \ref{subsec:cropping} (See figure ~\ref{fig:Fig1}-A). Then, we employ a pre-trained CNN network with frozen weights to extract so-called “deep features” from 72 maximum intensity projections computed from 5-degree steps of axial rotations of the cropped head and neck PET volumes, resulting in 72 feature vectors per each patient, as elaborated in detail in sections \ref{subsec:feat_ext} and \ref{subsec:conv_feat_agg}. Next, we aggregate the extracted feature vectors of all views (MIPs) and process them through an ML pipeline for outcome prediction (figure ~\ref{fig:Fig1}-B). All the steps are described in detail in the following sections.

\subsection{Automatic Cropping of the Head and Neck Region}
\label{subsec:cropping}

\begin{figure}[h]
	\centering    
    \includegraphics[clip,width=1\linewidth,trim=3.7cm 4.5cm 2.8cm 3.9cm]{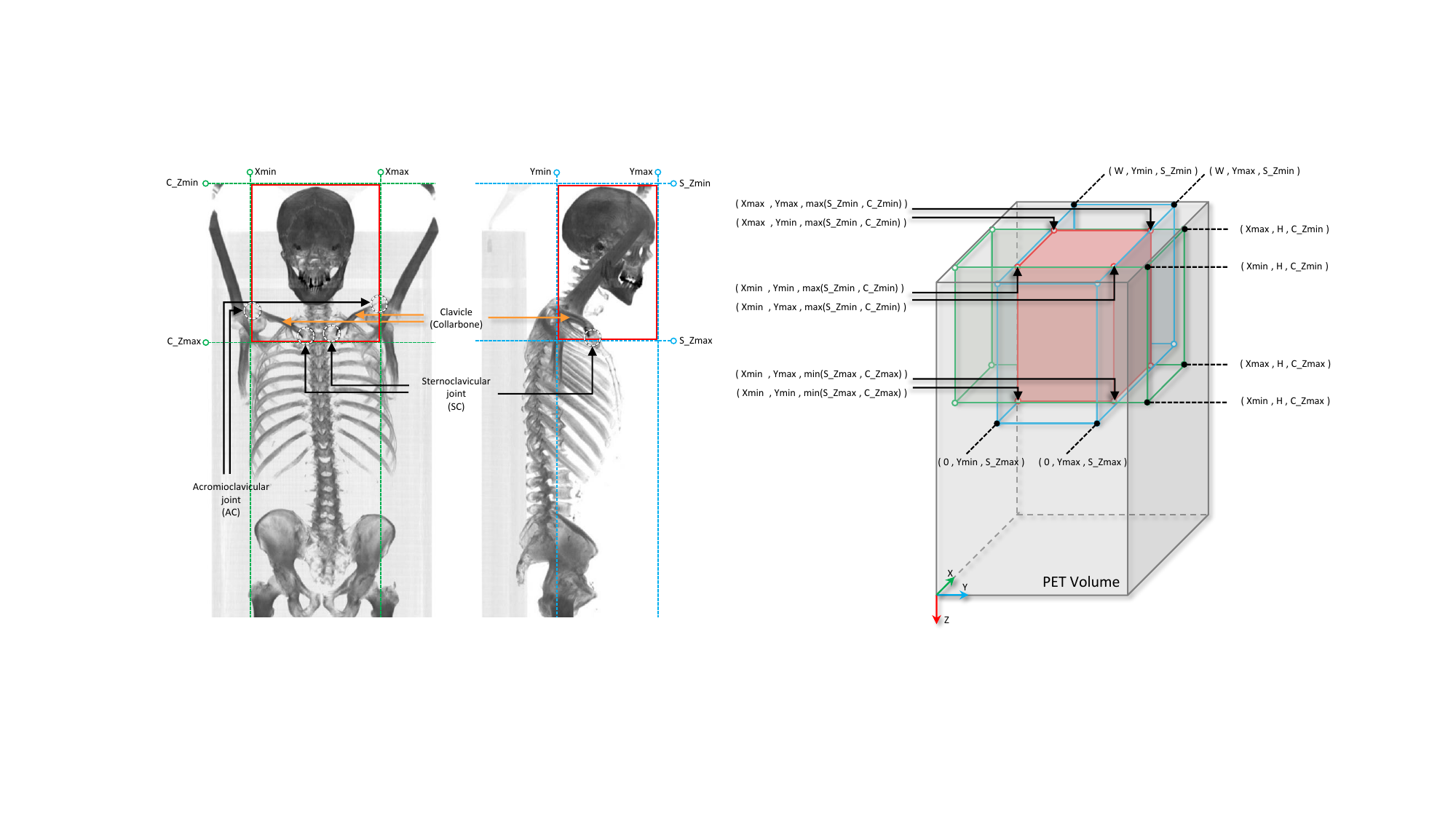}
\vspace{-1.0em}
	\caption{Visual summary of the proposed bounding box definition and cropping for locating the anatomical head and neck region on the PET volumes of the patients based on the coronal and Sagittal MIP CT projections. Locating the bounding boxes on coronal and sagittal CT MIPs based on the AC and SC clavicle joints (left), and the 3D bounding box cropping on the PET volume based on the two 2D bounding boxes on the CT mips (right).}
	\label{fig:Fig2}
\end{figure}

HNC tumors mainly affect the upper aerodigestive tract. in the early-stage disease the risk of distant metastasis is very low, where only approximately 10\% of patients have distant metastasis at the time of diagnosis \cite{melchardt2018clonal}. Given that, in order to focus on the main site of the disease affection, we employed a deep learning-based object detection model to automatically locate the anatomical upper aerodigestive region. To this end, we first generate a coronal and a sagittal maximum intensity projection of the CT scan of each patient for the whole dataset. Then, we manually annotate the anatomical head and neck region bounding boxes on both coronal and sagittal CT MIPs. On the coronal MIP, we placed a bounding box vertically between the sternoclavicular (SC) joint (the link between the clavicle or collarbone, and the sternum, or breastbone) and the top of the head, or the top of the image, if the full skull is not covered in the CT scan image. And horizontally between the left and right acromioclavicular (AC) joint (the joint where the scapula or shoulder blade, articulates with the clavicle or collarbone). (See figure ~\ref{fig:Fig2}). We target the clavicle bone to extend the oropharyngeal region down, mainly to contain all the primary tumors and the involved local lymph nodes in the anatomical region of interest. Since we defined the anatomical region of interest based on the regional bones, for generating coronal and sagittal CT MIPs, we clipped all CT voxel values in the corresponding range of hard tissue (bone) hounsfield units (700 - 2000). 

\subsection{Training the Object Detection Model on CT Projections}
\label{subsec:od_training}
For the object detection architecture we used TridentNet \cite{Li2019-ge}. TridentNet is a multi-stage deep learning-based 2D object detection model initially proposed for handling the highly variating object scales. Utilizing deformable convolution in its "Trident" blocks, TridentNet extracts multi-scale features with different receptive fields for different object scales from features of the same semantic layers. Based on our previous experience with this architecture for detecting very small and highly ambiguous objects in PET images \cite{toosi2023advanced, toosi2023state}, here we picked the same architecture for localizing the anatomical head and neck region in CT coronal and sagittal MIPs. To this end, first the training set was divided into 5 folds, 4 for training a network and one for testing. Both coronal and sagittal MIPs were fed to the networks for training, along with the manually made bounding boxes as ground truth. The predicted bounding boxes per each patient in 5 testing folds were then used for cropping the head and neck region on the corresponding PET volumes.

\subsection{PET Multi-Angle Maximum Intensity Projection (MA-MIP)}
\label{subsec:MA-MIP}
After cropping the head and neck region on the PET volume, the cropped regions underwent the feature extraction step.
The common practice for extracting features from biomedical volumetric images are either extracting features directly from the the 3D volume, or the pixels of the transaxial layers. But in our proposed method we instead extract image features from multiple maximum intensity projections (MA-MIPs) of the axial rotations of the cropped PET volume. To this end, we step by step rotate the cropped PET volume axially for 5 degrees, and per each rotation a MIP is captured from the rotated volume. This step is repeated 72 times in every 5 degrees to cover one complete axial rotation (see figure~\ref{fig:Fig1}, B). This enables us to leverage strong 2D convolutional neural networks backbones pretrained on large-scale public datasets like ImageNet, for extracting "deep" features.

\subsection{Deep Feature Extraction from MA-MIP Projections}
\label{subsec:feat_ext}
The vast size and diversity of ImageNet, with millions of natural images belonging to thousands of categories, have allowed training of models with robust image recognition capabilities that can effectively generalize across different visual domains, capturing complex patterns, textures, and structural relationships. 
By training a deep neural network on this dataset, the model learns to recognize features that are not specific to any particular class or modality, but rather generalize well across different visual domains. 

Through transfer learning, the knowledge gained from one task can be transferred to another related task. By doing so, the model can leverage the knowledge it has acquired and adapt it to the new task, leading to improved performance compared to training a model from scratch. When using a pre-trained CNN backbone for biomedical image analysis, the early layers can capture low-level features like texture, color, and shape, while higher layers can capture more abstract features like organs, tissues, or lesions. This hierarchical representation enables the model to capture both local and global information in medical images, leading to better performance in various tasks \cite{kora2022transfer}. 

In spite of the significant variability of medical images, mainly due to factors like patient demographics, acquisition protocols, and pathological conditions, the robustness of pre-trained CNN backbones on ImageNet can translate to medical images, helping the model to generalize across different populations, modalities, and diseases \cite{alzubaidi2021novel}. 

\subsubsection{EfficientNet as the feature extraction CNN backbone}
Looking at the trends of the proposed convolutional neural network architectures on the ImageNet \cite{deng2009imagenet}, from the emerge of AlexNet \cite{krizhevsky2012imagenet} in 2012 till present, the proposed CNN architectures are becoming larger in order to gain higher accuracy. However, the more the architectures get larger the more the number of parameters they reach, so much as it seems they have hit the hardware limits \cite{alzubaidi2021review}.

The most common approach for scaling the convolutional neural networks is increasing the depth, in order to help the network to learn more complex features throughout the depth and to generalize well on new tasks. This comes with the cost of difficulties in terms of training such deep models due to the well-known problems such as gradient vanishing \cite{tan2019vanishing}. Despite the proposed techniques to overcome this issue such as batch normalization and skip connections, the accuracy of deeper networks seems to saturate at certain point \cite{philipp2017exploding}.

Among the state-of-the-art architectures designed for ImageNet, EfficientNet \cite{tan2019efficientnet} is an attempt to achieve higher accuracy, by uniformly scaling the CNN architecture's depth, width, and resolution, forming a compound scaling framework. In order to scaling up the CNN network while keeping the number of parameters of the network low, as the main building block, EfficientNet use Inverted Bottleneck Block previously proposed in MobileNetV2 architecture \cite{sandler2018mobilenetv2}, along with squeeze and excitation optimization \cite{hu2018squeeze}. The essence of Inverted Bottleneck Block is using the Depthwise separable convolution layer instead of regular convolution block, which splits the convolution operation into two separate operation, a depth wise convolution and a pointwise convolution. This splitting lets the model to perform same convolution operation in each block with much less computation. First introduced in MobileNet architecture \cite{howard2017mobilenets}, the depthwise separable convolution enables the EfficientNet architecture to scale up and gain higher accuracy with much lower number of parameters. 

Building on top of a baseline model, called EfficientNetB0, seven scales of EfficientNet is provided with different uniform scales (EfficientNet B1 - B7) which could achieve higher accuracies compared to the state-of-the-art CNN architectures with same depth on the ImageNet dataset, while having lower parameters and being faster. Besides the superior performance on the ImageNet, EfficientNets also achieved state of-the-art accuracy on many other widely used datasets by transfer learning, while having up to 21x less parameters compared to other CNNs \cite{tan2019efficientnet}.

The remarkable performance of the EfficientNet architecture in terms of transfer learning, along with the uniform scale-up of the baseline architecture, motivated us to apply this model for feature extraction in our work. To this end, we selected 3 different scales of the EfficientNet pre-trained on ImageNet, namely B1, B4, and B7 (small, medium, and large) for deep features extraction from the MA-MIPs of the cropped PET volumes, to compare the effect of model scale on the effectiveness of the extracted features for the outcome prediction task. 

\subsection{Deep Features Preparation}
\label{subsec:conv_feat_agg}
The output of the pre-trained EfficientNet used for feature extraction is the last convolutional block before the classifier head of the network which is in shape of a 4D tensor (a batch of 3D tensors). In order to make a feature vector from the 3D feature tensor for each image in the batch, different methods are commonly used. The most common approach is to apply a global pooling layer on the output of the convolution block, either an average pooling or a max pooling, to summarize the feature tensor to a one dimensional feature vector. The global max pooling operation essentially computes the maximum value over the entire height and width of a 2D convolutional block's channel in the 3D tensor output's depth. The global max pooling operation can be formulated as follows: 
$$\text{GlobalMaxPool}(\mathbf{T})_{c} = \max_{h,w} (\mathbf{T}_{c,h,w})$$
Here, $\mathbf{T} \in \mathbb{R}^{C \times H \times W}$ represents the 3D tensor output of the 2D convolutional block, and the index $c$ corresponds to the channels dimension. The indices $h$ and $w$ iterate over the spatial dimensions of the tensor, and the $max$ operation finds the maximum value among all the spatial locations in each channel for every instance. In global average pooling, similar to global max pooling operation, the $mean$ among all the spatial locations in each channel is computed as following.
$$\text{GlobalAveragePool}(\mathbf{T})_{c} = \frac{1}{h \times w} \sum_{i=1}^{h} \sum_{j=1}^{w} \mathbf{T}_{c,i,j}$$
In this work we also extended the global pooling layer to median and also standard deviation pooling in order examine different possible global pooling operators.

\subsection{ MA-MIPs Extracted Deep Features Fusion}
\label{subsec:feat_fusion}
Features extracted from MA-MIPs taken from different rotations of the cropped PET volume, can be seen as the information extracted from different views of the PET volume. Therefore, combining the information extracted from different views could give a more informative set of information about the region. 

In order to aggregate the feature vectors extracted from 72 MA-MIPs from different views of the cropped PET volume, feature fusion methods are applied. Multiple feature fusion methods is used in this work in order to aggregate feature vectores from multiple views into a single feature vector. The utilized methods can be categorized into three main groups; Statistical summarizing methods, linear transformation methods, and non-linear transformation methods.

Statistical Summarizing methods leverage simple statistical measures to aggregate information from the 72 feature vectors into a single feature vector. To this end the index-wise (channel-wise) maximum, mean, median, and standard deviation of the feature vectors is computed after stacking all the 72 feature vectors of the different views. 
The maximum value represents the most prominent feature across all views, while the mean, median, and standard deviation capture different aspects of the overall statistical distribution of features from different views.
These methods simplify the representation of the data by distilling it down to key statistical characteristics.

Linear Transformation methods on the other hand, encompass dimensionality reduction techniques applied index-wise (channel-wise) to the stacked feature vectors. Initially, the 72 feature vectors are stacked vertically to form a matrix, similar to the statistical summarizing methods, where each row represents a feature from one view, and there are 72 columns, each corresponding to a specific feature index (channel). Independent Component Analysis (ICA) is then individually applied to each column of the stacked feature vectors. This process summarizes the 72 cells within each feature index into a single value. By applying these linear transformations to each index separately, a reduced-dimensional representations for each feature index is obtained, in order to capture essential patterns and reduce dimensionality, while preserving unique characteristics specific to each view. This approach analyses how each feature contributes to the fused representation.

As for non-linear transformation methods we used an auto-encoder. This learning-based method is applied on the concatenated 72 feature vectors to learn a compact, lower-dimensional representation in an unsupervised end-to-end manner. The auto-encoder comprises an encoder network that maps the input data into a reduced-dimensional latent space and a decoder network that reconstructs the input from this latent representation. By training the auto-encoder, it learns to capture the most salient features in the data while discarding noise and redundancies, resulting in an optimized fused feature vector that retains essential information. These three categories of fusion methods represent a well-rounded approach to fusing feature vectors from multiple views of the HNC PET images. Each category offers unique advantages and insights into the data, allowing for a comprehensive analysis that may enhance the predictive power of the proposed model for outcome prediction.

\subsection{ML Prediction Pipeline for Recurrence-Free Survival Analysis}
\label{subsec:ML_pipeline}

Fused multi-view features are then fed to an outcome prediction pipeline in order to perform recurrence-free survival prediction.
Feature vectors of all 489 patients fist undergone a standard scaling by removing the mean and scaling to unit variance. A nested 5-fold cross validation with 20 times repetition is applied to evaluate the outcome prediction pipeline. Per each inner fold, highly correlated features were eliminated, then an independent component analysis (ICA) were applied on the remaining features to further reduce the dimensionality of the features also to avoid over-fitting. Cox proportional hazard survival  method were employed on the reduced dimension features to predict the survival risk of each patient. Grid search was used for parameter tuning. 
The trained models were tested on each of the outer unseen testing folds and the performance of the trained outcome prediction models are reported in terms of the mean and the standard deviation of the concordance index (c-index) of the five unseen test folds. In the next section we report the results of our proposed pipeline for recurrence-free survival analysis, and evaluate the effect of using different methods for each step throughout the proposed segmentation-free outcome prediction pipeline.

\section{Results}
\label{sec:results}

In this section we report the result of our proposed segmentation-free outcome prediction model for recurrence-free survival prediction of HNC patients using the multi-angle maximum intensity projections of their FDG PET images. 
Table \ref{tab:table_a} summarizes the result of all possible combinations of the feature extraction backbones, pooling methods, and deep feature fusion methods, over all 5 test folds, in terms of mean and standard deviation c-index. Rows of the Table \ref{tab:table_a} are divided into four sections each corresponding to one of four global pooling methods. On average over all three different sizes of the feature extraction method (EfficientNet B1, B4, and B7), and using all six different feature fusion methods, global average pooling seems to be slightly more effective compared to other pooling methods, with the mean c-index of $0.659$ compared to $0.653$, $0.645$, and $0.630$ for standard deviation, median, and max pooling, respectively. However, the difference is minimal. 

Comparing feature fusion methods, independent component analysis (ICA) method seems to work poorly with respect to other fusion methods ($0.519$ on average over four pooling methods and three backbone sizes). Among two other fusion methods, on average, statistical summarization methods seems to outperform non-linear transformation method (auto-encoder) with mean value over all five channel-wise methods equal to $0.674$ compared to $0.668$ for the auto-encoder. Between the four statistical summarization fusion methods, channel-wise maximum fusion with the mean c-index of 0.686 outperforms other methods with mean c-index of $0.673$, $0.672$, and $0.668$ for channel-wise standard deviation, mean, and median, respectively. 
In order to compare the effect of using the EfficientNet as feature extraction backbone with different sizes, looking at the average of the c-index values over all the proposed pooling and fusion methods, EfficientNet Large (B7) and EfficientNet Small (B1) with mean c-index of $0.658$ and $0.654$ respectively, slightly outperforming the EfficientNet Medium (B4) with the mean c-index of $0.630$, with the large size performing minimally better than EfficienNet Small.

\begin{table}[H]
\footnotesize
\medskip
    \centering
    \begin{tabular}{l|cccc|c|c} 
         Feature Extractor        & Mean                  & Median               & Max                    & STD-Dev               & ICA                   & AutoEncoder          \\ \hline
        \tiny{Global Average Pooling} &                   &                       &                       &                       &                       &                      \\ 
         EfficientNet Small (B1)  & $0.706\pm$\tiny{0.05} & $0.707\pm$\tiny{0.05} & $0.707\pm$\tiny{0.05} & $0.678\pm$\tiny{0.04} & $0.593\pm$\tiny{0.06} & $0.689\pm$\tiny{0.08}\\ 
         EfficientNet Medium (B4) & $0.615\pm$\tiny{0.03} & $0.638\pm$\tiny{0.03} & $0.626\pm$\tiny{0.05} & $0.640\pm$\tiny{0.04} & $0.590\pm$\tiny{0.08} & $0.644\pm$\tiny{0.05}\\ 
         EfficientNet Large  (B7) & $0.718\pm$\tiny{0.03} & $0.710\pm$\tiny{0.03} & $0.719\pm$\tiny{0.03} & $0.705\pm$\tiny{0.04} & $0.455\pm$\tiny{0.10} & $0.721\pm$\tiny{0.03}\\ 
                                  &                       &                       &                       &                       &                       &                      \\ 

        \tiny{Global Max Pooling} &                       &                       &                       &                       &                       &                      \\ 
         EfficientNet Small (B1)  & $0.613\pm$\tiny{0.08} & $0.613\pm$\tiny{0.07} & $0.703\pm$\tiny{0.06} & $0.680\pm$\tiny{0.03} & $0.541\pm$\tiny{0.03} & $0.639\pm$\tiny{0.06}\\ 
         EfficientNet Medium (B4) & $0.650\pm$\tiny{0.07} & $0.633\pm$\tiny{0.06} & $0.667\pm$\tiny{0.04} & $0.646\pm$\tiny{0.04} & $0.576\pm$\tiny{0.07} & $0.673\pm$\tiny{0.05}\\ 
         EfficientNet Large  (B7) & $0.625\pm$\tiny{0.03} & $0.634\pm$\tiny{0.03} & $0.697\pm$\tiny{0.05} & $0.665\pm$\tiny{0.04} & $0.459\pm$\tiny{0.03} & $0.651\pm$\tiny{0.03}\\
                                  &                       &                       &                       &                       &                       &                      \\ 
                                  
        \tiny{Global Median Pooling} &                    &                       &                       &                       &                       &                      \\ 
         EfficientNet Small (B1)  & $0.684\pm$\tiny{0.06} & $0.666\pm$\tiny{0.07} & $0.679\pm$\tiny{0.04} & $0.666\pm$\tiny{0.04} & $0.501\pm$\tiny{0.08} & $0.671\pm$\tiny{0.05}\\ 
         EfficientNet Medium (B4) & $0.652\pm$\tiny{0.02} & $0.663\pm$\tiny{0.03} & $0.651\pm$\tiny{0.04} & $0.641\pm$\tiny{0.05} & $0.491\pm$\tiny{0.03} & $0.630\pm$\tiny{0.06}\\ 
         EfficientNet Large  (B7) & $0.688\pm$\tiny{0.06} & $0.678\pm$\tiny{0.05} & $0.701\pm$\tiny{0.04} & $0.697\pm$\tiny{0.04} & $0.556\pm$\tiny{0.07} & $0.698\pm$\tiny{0.06}\\
                                  &                       &                       &                       &                       &                       &                      \\ 

        \tiny{Global STD-Dev Pooling} &                   &                       &                       &                       &                       &                      \\ 
         EfficientNet Small (B1)  & $0.702\pm$\tiny{0.05} & $0.701\pm$\tiny{0.06} & $0.701\pm$\tiny{0.06} & $0.709\pm$\tiny{0.02} & $0.456\pm$\tiny{0.05} & $0.685\pm$\tiny{0.08}\\ 
         EfficientNet Medium (B4) & $0.674\pm$\tiny{0.04} & $0.655\pm$\tiny{0.03} & $0.669\pm$\tiny{0.04} & $0.655\pm$\tiny{0.05} & $0.503\pm$\tiny{0.03} & $0.649\pm$\tiny{0.03}\\ 
         EfficientNet Large  (B7) & $0.708\pm$\tiny{0.04} & $0.706\pm$\tiny{0.04} & $0.713\pm$\tiny{0.04} & $0.689\pm$\tiny{0.06} & $0.506\pm$\tiny{0.04} & $0.665\pm$\tiny{0.04}\\ 
         
    \end{tabular}
    \vspace{0.5cm}
    \caption{ This table summarizes the result of all combinations of feature extraction CNN backbones, pooling methods, and deep feature fusion methods, over all 5 test folds, in terms of mean and standard deviation c-index. Rows of the table are divided into four sections, each corresponding to one of four global pooling methods. Columns are also divided into three separate sections each corresponding to one fusion method, from left to right channel-wise statistical summarization methods, linear transformation method, and non-linear transformation method.}
    \label{tab:table_a}
\end{table}

\begin{figure}[h]
	\centering    
    \includegraphics[clip,width=1\linewidth,trim=0.0cm 0.5cm 7.5cm 0.0cm]{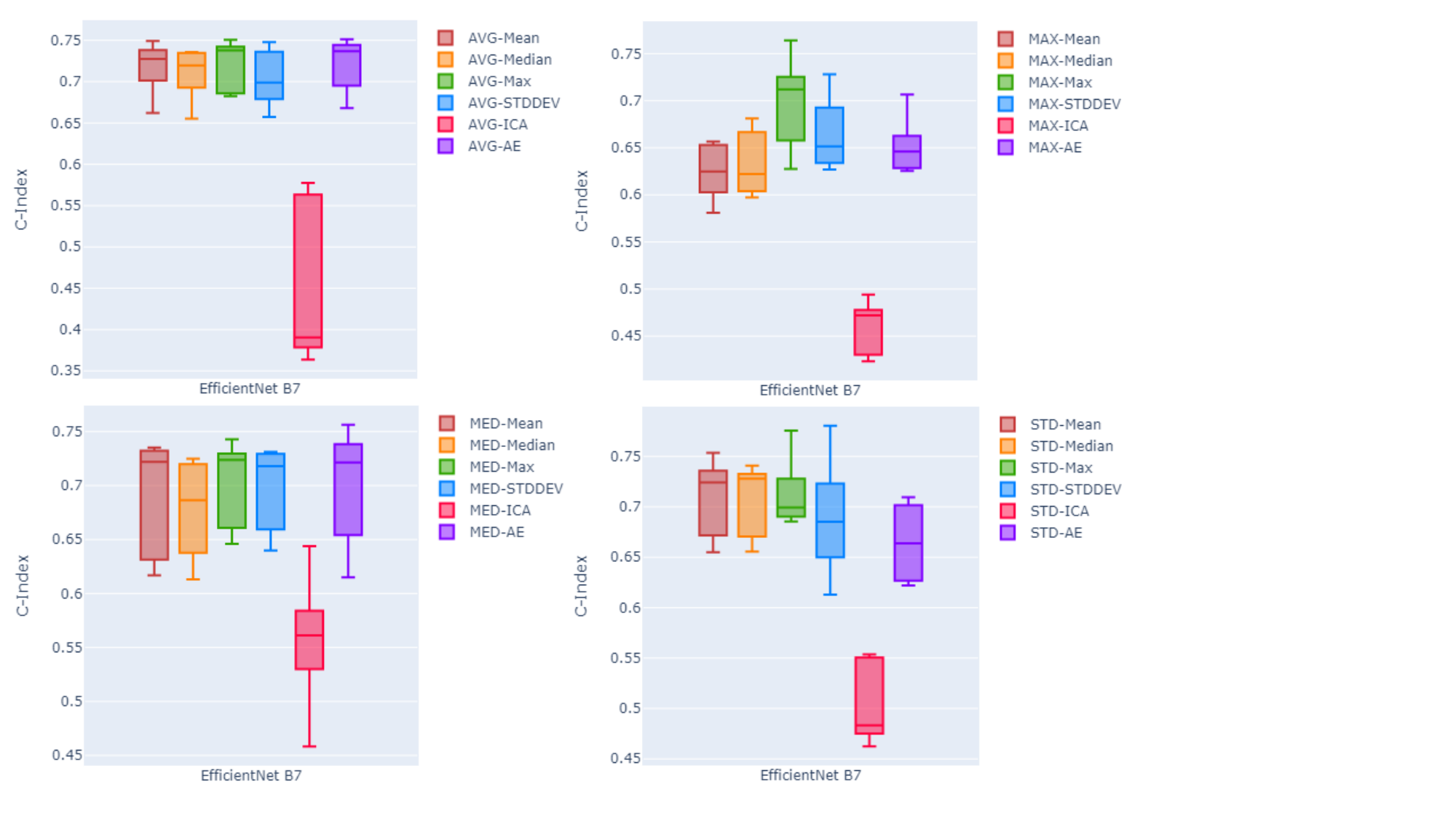}
\vspace{-1.0em}
	\caption{Box plots of the c-index values over all five test folds. Each plot shows the results using one of the four proposed pooling methods, for EfficientNet large as the feature extraction method, and all six different fusion techniques. Global average pooling, top left. Global max pooling top right. Global median pooling, buttom left, and global standard deviation pooling on the buttom right.}
	\label{fig:poolings}
\end{figure}

Picking the EfficienNet Large as the better performing feature extraction method, in order to compare the effectiveness of the proposed pooling methods, figure \ref{fig:poolings} shows box plots of the c-index values of five test folds. Each plot shows the results using one of the four proposed pooling methods, for EfficientNet large, and all six different fusion techniques. Looking at the median and the variability of the box plots, channel-wise maximum fusion method seems performing slightly better than other statistical summarization methods and the auto-encoder method. However, for the global average pooling method, as the best pooling method among all proposed poolings, auto-encoder-based fusion seems to perform on par with the channel-wise maximum and mean fusion techniques. This is not true in other three diagrams showing results of methods using other pooling techniques. On the other hand, looking at the four statistical summarization methods and the non-linear transformation method (auto-encoder) in all four diagrams of the figure \ref{fig:poolings}, it seems that boxes are more compact with relatively higher median c-index values and lower variability compared to the boxes in other three diagrams corresponding to other pooling methods. 

\begin{figure}[h]
	\centering    
    \includegraphics[clip,width=1\linewidth,trim=1.5cm 4.5cm 2.0cm 4.5cm]{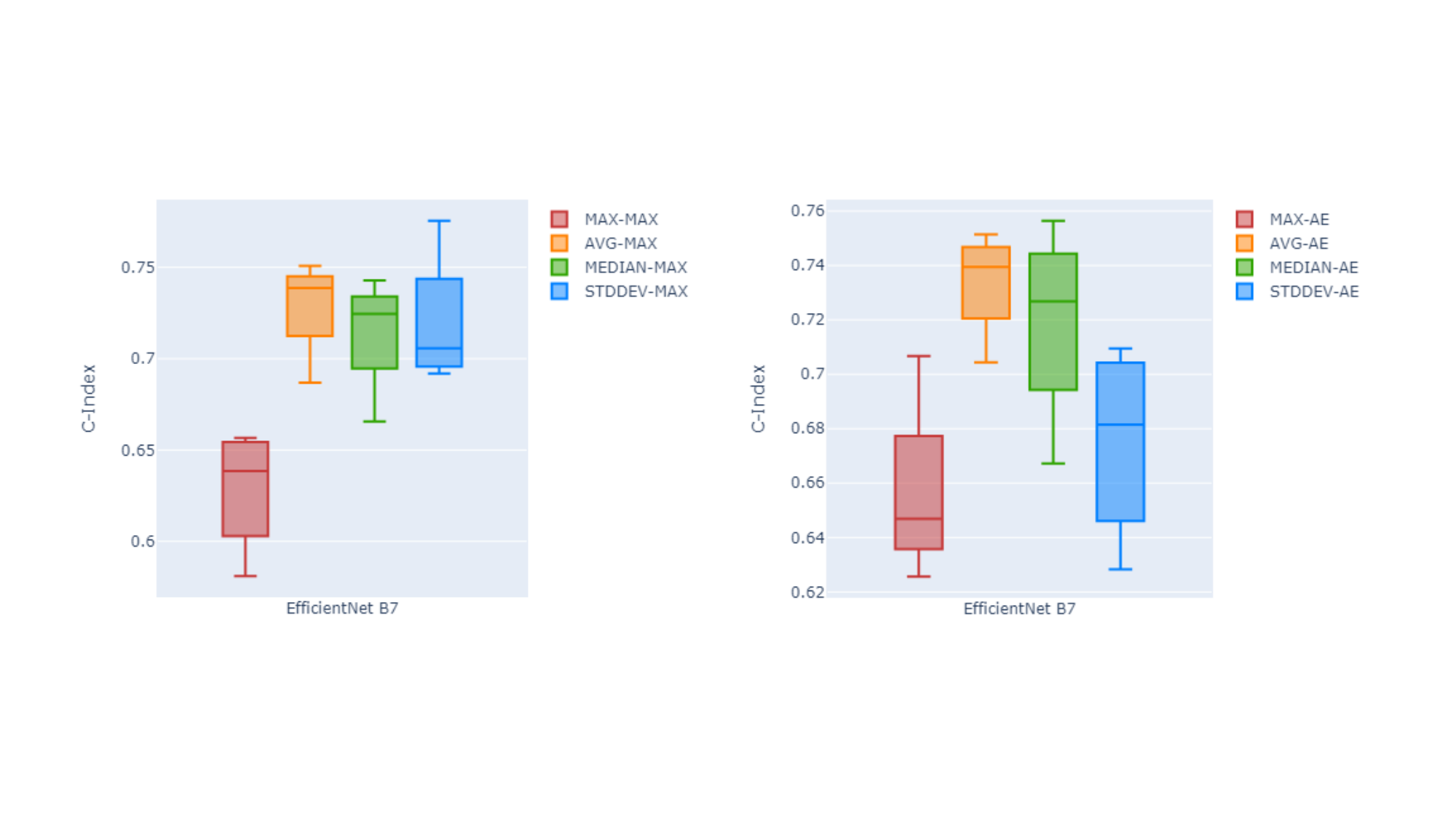}
\vspace{-1.0em}
	\caption{Box plots summarizing the performance of channel-wise maximum fusion technique(left) and the auto-encoder(right) as the two best fusion methods over all four different global pooling methods, and all five testing folds.}
	\label{fig:fusions}
\end{figure}

Likewise, figure \ref{fig:fusions} shows in box plots the performance of channel-wise maximum fusion technique and the auto-encoder as the two best fusion methods over all four different global pooling methods, over all five testing folds. Here as can be seen, global average pooling has higher median c-index and lower variability compared to other three pooling methods on both channel-wise maximum fusion and the auto-encoder.

\begin{table}[h]
\footnotesize
\medskip
    \centering
    \begin{tabular}{l|c|c|c} 
        Outcome Prediction Method                           & Mean c-index    & Max     & Min     \\ \hline
         L. Rebaud et al. \cite{rebaud2022simplicity}          & $0.688$ & $0.732$ & $0.642$ \\ 
         EfficientNet Large + GAvgPool + CW-Mean (Ours)     & $0.718$ & $0.749$ & $0.662$ \\ 
         EfficientNet Large + GAvgPool + CW-Median (Ours)   & $0.710$ & $0.736$ & $0.655$ \\ 
         EfficientNet Large + GAvgPool + CW-Max (Ours)      & $0.719$ & $0.751$ & $0.682$ \\ 
         EfficientNet Large + GAvgPool + CW-STD-Dev (Ours)  & $0.705$ & $0.748$ & $0.657$ \\ 
         EfficientNet Large + GAvgPool + AutoEncoder (Ours) & $0.721$ & $0.742$ & $0.668$ \\ 
    \end{tabular}
    \vspace{0.5cm}
    \caption{Comparison of the results of our proposed method, using global average pooling and EfficientNet large as feature extractor, to the baseline method \cite{rebaud2022simplicity} winner of the HECKTOR 2022 challenge for recurrence-free survival analysis task, on the training part of the dataset, in terms of mean, maximum, and minimum c-index over the test folds.}
    \label{tab:table_b}
\end{table}

Table \ref{tab:table_b} shows the results of the method proposed in \cite{rebaud2022simplicity} who won the best place in the HECKTOR 2022 challenge leaderboard for recurrence-free survival analysis task, on the training part of the dataset, in terms of mean, maximum, and minimum c-index over the test folds. Using global average pooling over the deep features extracted using EfficientNet with large size, and all statistical summarization and non-linear transformation fusion methods outperforms the result reported by \cite{rebaud2022simplicity} on the dataset.

\begin{table}[h]
\footnotesize
\medskip
    \centering
    \begin{tabular}{l|cccc|c|c} 
         Feature Extractor                                   & c-index               \\ \hline 
         InceptionResNetV2 \tiny{+ GAvgPool + CW-Max}        & $0.641\pm$\tiny{0.06} \\
         InceptionV3 \tiny{+ GAvgPool + CW-Max}              & $0.649\pm$\tiny{0.05} \\
         Resnet-152 \tiny{+ GAvgPool + CW-Max}               & $0.662\pm$\tiny{0.06} \\
         VGG16 \tiny{+ GAvgPool + CW-Max}                    & $0.640\pm$\tiny{0.07} \\
         VGG19 \tiny{+ GAvgPool + CW-Max}                    & $0.635\pm$\tiny{0.03} \\
         Xception \tiny{+ GAvgPool + CW-Max}                 & $0.615\pm$\tiny{0.09} \\
         DenseNet-201 \tiny{+ GAvgPool + CW-Max}             & $0.663\pm$\tiny{0.06} \\
         NASNet-Large \tiny{+ GAvgPool + CW-Max}             & $0.644\pm$\tiny{0.04} \\
         ConvNeXt-Base \tiny{+ GAvgPool + CW-Max}            & $0.660\pm$\tiny{0.06} \\ \hline
         EfficientNet Small (B1) \tiny{+ GAvgPool + CW-Max}  & $0.707\pm$\tiny{0.05} \\
         EfficientNet Medium (B4) \tiny{+ GAvgPool + CW-Max} & $0.626\pm$\tiny{0.05} \\
         EfficientNet Large  (B7) \tiny{+ GAvgPool + CW-Max} & $0.719\pm$\tiny{0.03} \\
         
    \end{tabular}
    \vspace{0.5cm}
    \caption{Result of using other architectures as feature extractors compared to EfficientNet in terms of mean c-index over the five test folds.}
    \label{tab:table_c}
\end{table}

We have also tried a few other well-known pre-trained CNN backbones for deep feature extraction in order to compare to the EfficientNet used in our proposed method. Table \ref{tab:table_c} summarizes the results of using 9 other pre-trained CNN backbones in terms of mean c-index over the five test folds. For all these feature extractors we used global average pooling and then channel-wise maximum fusion method for the PFS prediction pipeline. While EfficientNet with small and large sizes seem to outperform all the other feature extractors, other backbones are performing better or on par with EfficientNet medium. We report the result of all used CNN backbones, using all the proposed pooling methods and feature fusion methods in Table \ref{tab:table_d}.

\section{Discussion}
\label{sec:discussion}
The present study proposed a segmentation-free outcome prediction method for HNC patients. Instead of relying on manual segmentations, this work utilizes an object detection method that is trained to find the anatomical head and neck region of the patient on both coronal and sagittal MIPs of the patient's CT scan. The benefit of using such method for automatically cropping the anatomical head and neck region is that, compared to the time and resource demanding task of manual delineation of all the tumors and involved lymph nodes, detecting the anatomical site of the primary disease is a less time and resource demanding task. Defining the manual bounding boxes used in training the object detection method does not demand an expert physician. Constraining the entire space of the patient's body to only the primary site of the disease has also the advantage of helping the feature extraction methods to focus more on the relevant information, in such a way that no delineation of the disease related uptakes on the FDG PET will be needed. 

This point also can be seen in previous works on the second round of MICCAI's HECKTOR challenge, where all the DICOM volumes were accompanied with 3D bounding boxes with size of ${144 \times 144 \times 144}$ voxels containing the extended oropharyngeal region of the patients. The provided bounding boxes not only facilitated the use of 3D methods for automatic tumor segmentation due to the high computation and memory demands of such methods, but also made room for proposing outcome prediction methods extracting information directly from the given bounding box, instead of relying on the segmentation masks of the primary tumors, which resulted in the first and second best performing methods for the outcome prediction task \cite{Saeed2022-vu, Naser2022-nm} outperforming methods relying on predicting the segmentation masks of the primary tumors. It is worth noting that thanks to the recent success of methods such as TotalSegmentator \cite{wasserthal2023totalsegmentator}, training an object detection network to crop the region is not necessarily needed. Instead, targeting the bones used in this work for defining the anatomical head and neck site (section \ref{subsec:cropping}) one can use the segmentation masks of TotalSegmentator and use it to crop the ROI.

Beside using a bounding box of the entire disease site for extracting information (rather than manual delineations of the tumors/lymph nodes), in the present study, instead of extracting image features from the voxels of the 3D volumes of interest, or pixels of trans-axial slices of the 3D volume, feature extraction was performed on multiple maximum intensity projections of the VOI from different trans-axial angels. MIP projections from multiple axial angles can potentially enhance the representation of the malignancies (tumor and involved lymph nodes) and capture a broader view of its metabolic activity, while these information could be lost in common radiomics workflows. We have experienced the effect of enhanced representation of the malignancies in  previous works for detection and segmentation of small metastatic lesions of biochemically recurrent prostate cancer \cite{toosi2023state, toosi2023advanced}. The present work follows the idea of using multi-angle PET MIPs in cancer diagnosis tasks, but here specifically for the task of outcome prediction. 

MIPs have been used in the literature for the task of outcome prediction. FDG PET coronal and sagittal MIPs were used in \cite{girum202218f} for calculating two surrogate features to be replaced by their 3D equivalents, TMTV and Dmax, which were shown to be highly predictive in lymphoma cancer management. In \cite{ferrandez2023artificial}, a CNN network was trained on MIP projections of FDG PET scans to predict the probability of 2-year time-to-progression (TTP). However both mentioned works used only the coronal and sagittal MIPs compared to 72 multi-angle MIPs in our work. In \cite{girum202218f}, MIPs were used indirectly as a mean for estimating volumetric-based biomarkers through defining surrogate biomarkers, while our proposed method is using the information extracted from the MA-MIPs directly for the task of outcome prediction. Moreover, efforts in both \cite{ferrandez2023artificial} and \cite{girum202218f} were either explicitly or implicitly aware of tumor segmentation masks, while our proposed method is completely segmentation-free. 

Beyond the aim of our work in proposing a segmentation-free approach for HNC outcome prediction, there is a body of research on segmentation-free methods but for other aims. For instance, \cite{taghanaki2018segmentation} proposed a segmentation-free method for directly estimating tumor volume and  total amount of metabolic activity from patient PET images, which are two important imaging biomarkers in disease management and survival analysis of the patients. However, our present work bypasses the step of estimating the aforementioned biomarkers and directly predicts the outcome of the patients.

One of the main benefits of the proposed method in this work is the use of pre-trained CNN backbones with frozen weights for feature extraction. This property has multiple benefits. Firstly, using the pre-trained CNN backbones with frozen weights and without fine-tuning on the target dataset, makes the outcome prediction pipeline highly reproducible. This property is perfectly aligned with  efforts such as the Image Biomarker Standardization Initiative (IBSI) for standardising the extraction of image biomarkers for high-throughput quantitative image analysis (radiomics) \cite{zwanenburg2020image}. Moreover, it can also be beneficial in terms of computational resources. 

Using the pre-trained CNN backbones as a pure transfer learning method, without the need for training on the target domain dataset, will make the outcome prediction pipeline highly portable, without need of high-end computational resources like GPUs. This helps the proposed methods be more generalizeable from limited training data and thus applicable in the clinic, using limited computational resources or embedded/edge devices. In any case, we plan to also study fine-tuning of the CNN backbones before feature extraction to evaluate the effect of domain adaptation on the target dataset on the task of outcome prediction.

\begin{table}[H]
\footnotesize
\medskip
\centering

\begin{tabular}{l | cccccc}  

                             &\scriptsize{CW-Mean Fusion}&\scriptsize{CW-Median Fusion}&\scriptsize{CW-MAX Fusion}&\scriptsize{CW-STD-Dev Fusion}&\scriptsize{AutoEncoder Fusion}\\ \hline
\tiny{Global Max Pooling}    &                        &                        &                        &                        &                        \\
InceptionResNetV2            & $0.604\pm$\tiny{0.071} & $0.613\pm$\tiny{0.047} & $0.549\pm$\tiny{0.061} & $0.559\pm$\tiny{0.070} & $0.630\pm$\tiny{0.071} \\
InceptionV3                  & $0.640\pm$\tiny{0.056} & $0.627\pm$\tiny{0.041} & $0.643\pm$\tiny{0.049} & $0.638\pm$\tiny{0.037} & $0.637\pm$\tiny{0.050} \\
Resnet                       & $0.672\pm$\tiny{0.064} & $0.673\pm$\tiny{0.058} & $0.654\pm$\tiny{0.058} & $0.666\pm$\tiny{0.056} & $0.675\pm$\tiny{0.062} \\
VGG16                        & $0.644\pm$\tiny{0.051} & $0.644\pm$\tiny{0.044} & $0.686\pm$\tiny{0.032} & $0.631\pm$\tiny{0.066} & $0.599\pm$\tiny{0.026} \\
VGG19                        & $0.637\pm$\tiny{0.027} & $0.624\pm$\tiny{0.042} & $0.625\pm$\tiny{0.023} & $0.625\pm$\tiny{0.054} & $0.638\pm$\tiny{0.045} \\
Xception                     & $0.613\pm$\tiny{0.064} & $0.607\pm$\tiny{0.068} & $0.608\pm$\tiny{0.051} & $0.615\pm$\tiny{0.054} & $0.618\pm$\tiny{0.071} \\
DenseNet201                  & $0.685\pm$\tiny{0.060} & $0.679\pm$\tiny{0.053} & $0.689\pm$\tiny{0.054} & $0.612\pm$\tiny{0.037} & $0.683\pm$\tiny{0.026} \\
EfficientNetB1               & $0.613\pm$\tiny{0.083} & $0.613\pm$\tiny{0.068} & $0.703\pm$\tiny{0.056} & $0.680\pm$\tiny{0.033} & $0.639\pm$\tiny{0.056} \\
EfficientNetB4               & $0.650\pm$\tiny{0.068} & $0.633\pm$\tiny{0.055} & $0.667\pm$\tiny{0.036} & $0.646\pm$\tiny{0.044} & $0.673\pm$\tiny{0.053} \\
EfficientNetB7               & $0.625\pm$\tiny{0.031} & $0.634\pm$\tiny{0.032} & $0.697\pm$\tiny{0.052} & $0.665\pm$\tiny{0.041} & $0.651\pm$\tiny{0.033} \\
                             &                        &                        &                        &                        &                        \\

\tiny{Global Average Pooling}&                        &                        &                        &                        &                        \\
InceptionResNetV2            & $0.605\pm$\tiny{0.066} & $0.612\pm$\tiny{0.065} & $0.641\pm$\tiny{0.059} & $0.627\pm$\tiny{0.070} & $0.612\pm$\tiny{0.078} \\
InceptionV3                  & $0.635\pm$\tiny{0.052} & $0.630\pm$\tiny{0.049} & $0.649\pm$\tiny{0.048} & $0.654\pm$\tiny{0.043} & $0.651\pm$\tiny{0.050} \\
Resnet                       & $0.676\pm$\tiny{0.076} & $0.675\pm$\tiny{0.076} & $0.662\pm$\tiny{0.055} & $0.658\pm$\tiny{0.053} & $0.680\pm$\tiny{0.077} \\
VGG16                        & $0.641\pm$\tiny{0.052} & $0.624\pm$\tiny{0.056} & $0.640\pm$\tiny{0.072} & $0.614\pm$\tiny{0.092} & $0.655\pm$\tiny{0.061} \\
VGG19                        & $0.641\pm$\tiny{0.053} & $0.646\pm$\tiny{0.080} & $0.635\pm$\tiny{0.034} & $0.652\pm$\tiny{0.046} & $0.636\pm$\tiny{0.047} \\
Xception                     & $0.614\pm$\tiny{0.080} & $0.593\pm$\tiny{0.081} & $0.615\pm$\tiny{0.088} & $0.614\pm$\tiny{0.080} & $0.640\pm$\tiny{0.077} \\
DenseNet201                  & $0.668\pm$\tiny{0.061} & $0.665\pm$\tiny{0.066} & $0.663\pm$\tiny{0.064} & $0.653\pm$\tiny{0.055} & $0.683\pm$\tiny{0.074} \\
EfficientNetB1               & $0.706\pm$\tiny{0.048} & $0.707\pm$\tiny{0.051} & $0.707\pm$\tiny{0.051} & $0.678\pm$\tiny{0.035} & $0.689\pm$\tiny{0.076} \\
EfficientNetB4               & $0.615\pm$\tiny{0.028} & $0.638\pm$\tiny{0.029} & $0.626\pm$\tiny{0.051} & $0.640\pm$\tiny{0.041} & $0.644\pm$\tiny{0.048} \\
EfficientNetB7               & $0.718\pm$\tiny{0.033} & $0.710\pm$\tiny{0.033} & $0.719\pm$\tiny{0.032} & $0.705\pm$\tiny{0.036} & $0.721\pm$\tiny{0.034} \\
                             &                        &                        &                        &                        &                        \\

\tiny{Global Median Pooling} &                        &                        &                        &                        &                        \\
InceptionResNetV2            & $0.628\pm$\tiny{0.062} & $0.636\pm$\tiny{0.063} & $0.635\pm$\tiny{0.033} & $0.641\pm$\tiny{0.048} & $0.630\pm$\tiny{0.045} \\
InceptionV3                  & $0.632\pm$\tiny{0.038} & $0.655\pm$\tiny{0.044} & $0.660\pm$\tiny{0.036} & $0.645\pm$\tiny{0.065} & $0.628\pm$\tiny{0.057} \\
Resnet                       & $0.613\pm$\tiny{0.070} & $0.625\pm$\tiny{0.100} & $0.645\pm$\tiny{0.093} & $0.633\pm$\tiny{0.090} & $0.596\pm$\tiny{0.078} \\
VGG16                        & $0.571\pm$\tiny{0.038} & $0.574\pm$\tiny{0.025} & $0.557\pm$\tiny{0.041} & $0.556\pm$\tiny{0.049} & $0.582\pm$\tiny{0.055} \\
VGG19                        & $0.524\pm$\tiny{0.091} & $0.531\pm$\tiny{0.056} & $0.544\pm$\tiny{0.069} & $0.553\pm$\tiny{0.101} & $0.561\pm$\tiny{0.061} \\
Xception                     & $0.579\pm$\tiny{0.073} & $0.595\pm$\tiny{0.080} & $0.612\pm$\tiny{0.095} & $0.606\pm$\tiny{0.092} & $0.613\pm$\tiny{0.058} \\
DenseNet201                  & $0.644\pm$\tiny{0.066} & $0.653\pm$\tiny{0.068} & $0.634\pm$\tiny{0.088} & $0.618\pm$\tiny{0.065} & $0.665\pm$\tiny{0.061} \\
EfficientNetB1               & $0.684\pm$\tiny{0.061} & $0.666\pm$\tiny{0.056} & $0.679\pm$\tiny{0.038} & $0.666\pm$\tiny{0.041} & $0.671\pm$\tiny{0.051} \\
EfficientNetB4               & $0.652\pm$\tiny{0.017} & $0.663\pm$\tiny{0.027} & $0.651\pm$\tiny{0.040} & $0.641\pm$\tiny{0.045} & $0.630\pm$\tiny{0.055} \\
EfficientNetB7               & $0.688\pm$\tiny{0.057} & $0.678\pm$\tiny{0.048} & $0.701\pm$\tiny{0.042} & $0.697\pm$\tiny{0.041} & $0.698\pm$\tiny{0.057} \\
                             &                        &                        &                        &                        &                        \\
                             
\tiny{Global STD-Dev Pooling}&                        &                        &                        &                        &                        \\ 
InceptionResNetV2            & $0.620\pm$\tiny{0.070} & $0.624\pm$\tiny{0.046} & $0.572\pm$\tiny{0.083} & $0.546\pm$\tiny{0.069} & $0.638\pm$\tiny{0.062} \\
InceptionV3                  & $0.634\pm$\tiny{0.634} & $0.635\pm$\tiny{0.058} & $0.643\pm$\tiny{0.040} & $0.638\pm$\tiny{0.040} & $0.639\pm$\tiny{0.066} \\
Resnet                       & $0.677\pm$\tiny{0.071} & $0.679\pm$\tiny{0.072} & $0.643\pm$\tiny{0.061} & $0.663\pm$\tiny{0.058} & $0.641\pm$\tiny{0.053} \\
VGG16                        & $0.654\pm$\tiny{0.046} & $0.648\pm$\tiny{0.035} & $0.661\pm$\tiny{0.064} & $0.630\pm$\tiny{0.051} & $0.628\pm$\tiny{0.042} \\
VGG19                        & $0.644\pm$\tiny{0.055} & $0.635\pm$\tiny{0.073} & $0.627\pm$\tiny{0.058} & $0.629\pm$\tiny{0.056} & $0.638\pm$\tiny{0.024} \\
Xception                     & $0.607\pm$\tiny{0.066} & $0.618\pm$\tiny{0.082} & $0.622\pm$\tiny{0.077} & $0.606\pm$\tiny{0.067} & $0.617\pm$\tiny{0.082} \\
DenseNet201                  & $0.670\pm$\tiny{0.049} & $0.664\pm$\tiny{0.032} & $0.657\pm$\tiny{0.045} & $0.655\pm$\tiny{0.031} & $0.682\pm$\tiny{0.045} \\
EfficientNetB1               & $0.702\pm$\tiny{0.051} & $0.701\pm$\tiny{0.058} & $0.701\pm$\tiny{0.058} & $0.709\pm$\tiny{0.015} & $0.685\pm$\tiny{0.075} \\
EfficientNetB4               & $0.674\pm$\tiny{0.041} & $0.655\pm$\tiny{0.032} & $0.669\pm$\tiny{0.044} & $0.655\pm$\tiny{0.047} & $0.649\pm$\tiny{0.026} \\
EfficientNetB7               & $0.708\pm$\tiny{0.041} & $0.706\pm$\tiny{0.038} & $0.713\pm$\tiny{0.036} & $0.689\pm$\tiny{0.062} & $0.665\pm$\tiny{0.040} \\

\end{tabular}
\vspace{0.5cm}
\caption{Complete result of all used pre-trained CNN backbones, using all the four proposed pooling methods and five feature fusion methods from two groups of channel-wise statistical summarization fusions and non-linear transformation fusion method in terms of mean and std-dev of c-index over all five test folds.}
\label{tab:table_d}
\end{table}

\section{Conclusion}

We successfully developed an easy-to-perform segmentation-free outcome prediction method, specifically for predicting  progression-free survival  of  head and neck cancer patients from their FDG PET-CT image volumes. We found that by using multiple rotation maximum intensity projections of the  PET images, and focusing the anatomical head and neck region guided by CT scan images of the patients,  outcome prediction is not only feasible, but is also able to outperform conventional methods, without the need of manual segmentation of the primary tumors or the involved/metastatic lymph nodes by the expert nuclear medicine physicians.

\section*{Acknowledgment}

This work was supported by the Canadian Institutes of Health Research (CIHR) Project Grant PJT-162216, Natural Sciences and Engineering Research Council of Canada (NSERC) Discovery Grant RGPIN-2019–06467, and through computational resources and services provided by Microsoft for Health.

\footnotesize
\RaggedRight
\bibliographystyle{unsrt}
\bibliography{biblio}
\end{document}